\documentclass[aps, prl, 
amsmath, showpacs, preprintnumbers, twocolumn, amssymb, amsmath, superscriptaddress]{revtex4-1}
\usepackage{graphicx}
\usepackage{mathptmx, textcomp}
\usepackage[latin1]{inputenc}
\usepackage{tabularx}
\usepackage{units}
\usepackage{color}
\usepackage{bm}

\begin{document}
\title{Electric feedback cooling of single charged nanoparticles in an optical trap}
\author{M.\,Iwasaki}
\affiliation{Department of Physics, Tokyo Institute of Technology, Ookayama 2-12-1, Meguro-ku, 152-8550 Tokyo}
\author{T.\,Yotsuya}
\affiliation{Department of Physics, Tokyo Institute of Technology, Ookayama 2-12-1, Meguro-ku, 152-8550 Tokyo}
\author{T.\,Naruki}
\affiliation{Department of Physics, Tokyo Institute of Technology, Ookayama 2-12-1, Meguro-ku, 152-8550 Tokyo}
\author{Y.\,Matsuda}
\affiliation{Department of Physics, Tokyo Institute of Technology, Ookayama 2-12-1, Meguro-ku, 152-8550 Tokyo}
\author{M.\,Yoneda}
\affiliation{Department of Physics, Tokyo Institute of Technology, Ookayama 2-12-1, Meguro-ku, 152-8550 Tokyo}
\author{K.\,Aikawa}
\affiliation{Department of Physics, Tokyo Institute of Technology, Ookayama 2-12-1, Meguro-ku, 152-8550 Tokyo}

\date{\today}


\begin{abstract}
We demonstrate feedback cooling of the center-of-mass motion of single charged nanoparticles to millikelvin temperatures in three dimensions via applying oscillating electric fields synchronized to their optically observed motion. The observed motional temperatures at weak feedback agree with a simple model and allow us to estimate the charge number of trapped nanoparticles. The agreement between our model and experiments is confirmed by independent measurements of the charge numbers based on a shift in the oscillation frequency induced by a constant electric field. The demonstrated temperature of below $\unit[10]{mK}$ at $\unit[4\times 10^{-3}]{Pa}$ is lower than that with the conventional optical cooling approach at this pressure by one to two orders of magnitude. Our results form the basis of manipulating cold charged nanoparticles and paves the way to quantum mechanical studies with trapped nanoparticles near their ground state.
\end{abstract}

\maketitle

%Introduction:\\
%\section{Introduction}
%\label{sec_intro}

%Intro
%\section{Introduction}
Manipulating the motion of objects near their quantum ground state has been a crucial subject in diverse fields from quantum simulations~\cite{lewenstein2007ultracold,bloch2008many,blatt2012quantum} and  quantum information processing~\cite{haffner2008quantum} to precision measurements~\cite{takamoto2005optical,rosenband2008frequency}. Cooling atomic ions and ensembles of neutral atoms to their motional ground state has been successful~\cite{bloch2008many,leibfried2003quantum}. Specific vibrational modes of nano- and micromechanical oscillators have been brought to their quantum ground state~\cite{chan2011laser,teufel2011sideband}. However, cooling the motion of particles including more than a few atoms to their motional ground state has been an elusive goal. The main difficulty lies in the absence of an efficient mechanism for cooling. 

Cold nanoparticles are expected to possess various applications such as testing quantum mechanics for macroscopic objects~\cite{romero-isart2011large,bassi2013models}, ultrasensitive force and mass sensing~\cite{chaste2012nanomechanical,yang2006zeptogram,stipe2001noncontact,moser2013ultrasensitive,ranjit2016zeptonewton,hempston2017force,hebestreit2018sensing}, and the laboratory test of the collisional dynamics of interstellar materials~\cite{shukla2009colloquium}. Up to now, cooling the motion of nanoparticles to millikelvin temperatures has been demonstrated via all-optical approaches, where trapping, observing, and cooling them are all based on light scattering~\cite{li2011millikelvin,gieseler2012subkelvin,kiesel2013cavity,millen2015cavity,vovrosh2017parametric,setter2018real}. The lowest temperature achieved with all-optical approaches is finally limited by random photon recoils~\cite{jain2016direct}. To overcome the limitation from photon recoils, an all-electrical approach for highly charged particles has been proposed~\cite{goldwater2018levitated}.

\begin{figure}[t]
\includegraphics[width=0.99\columnwidth] {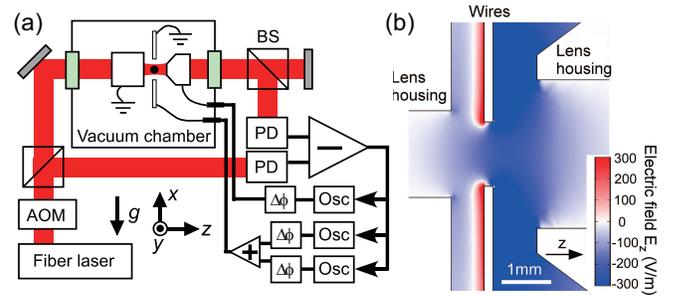}
\caption{(color online).  (a) Schematic representation of our experimental setup. Two objective lenses and two wires work as electrodes for applying three-dimensional electric fields around the trapping region. The two wires are tilted by 45$^\circ$ with respect to the $x$ direction such that they provide electric fields in both $x$ and $y$ directions. (b) Expanded schematic of the electrodes around the trap region on the plane including two wires. The numerically simulated electric field in the $z$ direction, with a dc voltage of $\unit[1]{V}$ applied on the right lens housing, is shown by colors.}
\label{fig:expset}
\end{figure}

Here, we show that the motional temperature of single charged nanoparticles in an optical trap is efficiently lowered via the optical measurement of the nanoparticle's position and the application of oscillating electric fields synchronized to their motion. The observed motional temperatures with the electric feedback $T_{\rm eff}$ agree with a simple model only when the mass of the nanoparticle is properly estimated through the time scale of the rethermalization of the motion after it is cooled. The agreement between our model and experimental results is confirmed by independent measurements of the charge number based on the electric-field-induced shift in the oscillation frequency. 

Compared to the conventional all-optical cooling method, parametric feedback cooling (PFC)~\cite{gieseler2012subkelvin,vovrosh2017parametric}, our method, electric feedback cooling (EFC), has two important advantages. First, while in PFC the cooling rate is proportional to $T_{\rm eff}$~\cite{gieseler2012subkelvin}, EFC has a high cooling rate determined by the magnitude of applied electric fields. Second, in EFC, the feedback signal is purely electrical, and thus does not perturb the optical position measurement, whereas, in PFC, the modulation on the trapping potential for cooling can affect the position measurement. In the present work, we demonstrate $T_{\rm eff}$ below $\unit[10]{mK}$ at $\unit[4\times10^{-3}]{Pa}$, which is about one to two orders of magnitude lower than the values obtained with PFC at this pressure, manifesting the efficiency of our approach. We estimate that, due to the high cooling rate of our approach, photon recoil heating will not be a major obstacle to cooling to near the quantum ground state at lower pressures. 

In our experiments, we trap silica nanoparticles with radii of about $\unit[100]{nm}$ in a one-dimensional optical lattice formed with a fiber laser at $\lambda=\unit[1550]{nm}$ [Fig.~\ref{fig:expset}(a)]~\cite{yoneda2017thermal,yoneda2018spontaneous,fnote7}. We observe the three-dimensional motion of nanoparticles with a photodetector measuring the spatio-temporal variation of the infrared light scattered by them~\cite{vovrosh2017parametric}. The area of the power spectral density (PSD) calculated from the photodetector signal is proportional to $T_{\rm eff}$~\cite{gieseler2012subkelvin,vovrosh2017parametric}. In most cases, trapped nanoparticles are positively ionized when the chamber is evacuated and an ion pressure gauge is turned on~\cite{frimmer2017controlling,fnote8}. If charging does not occur spontaneously, we induce charging with a corona discharge by applying about $\unit[400]{V}$ to an electrode placed near the trap region. 

\begin{figure}[t]
\includegraphics[width=0.99\columnwidth] {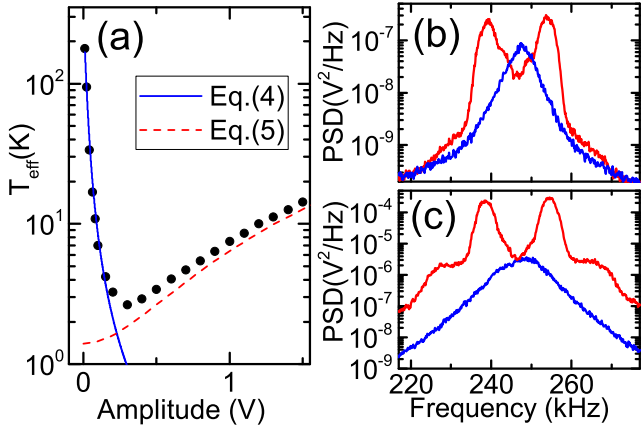}
\caption{(color online). (a) $T_{\rm eff}$ in the $z$ direction as a function of the applied voltage amplitude at $\unit[4.3]{Pa}$. (b) PSDs of the motion of trapped nanoparticles in the $z$ direction at small ($\unit[0.2]{V}$) and large ($\unit[1.5]{V}$) voltage amplitudes. (c) PSDs of the feedback signals for (b). }
\label{fig:tvsv}
\end{figure}

The motion of a trapped charged nanoparticle is continuously attenuated if an applied electric field switches its sign in phase with their motion and provides an electric force opposite to its velocity. Understanding on such a process is achieved by considering the following model. We describe the motion of nanoparticles in a specific direction in the presence of an electric field by the one-dimensional equation of motion:
\begin{equation}
\label{eq:eom}
\ddot{q}+(\Gamma_0+\Gamma_{\rm c})\dot{q}+\dot{q_n}+\Omega_0^2q=\dfrac{F_{\rm fl}}{m}
\end{equation}
where $q$, $q_n$, $\Gamma_0$, $\Gamma_{\rm c}$, $\Omega_0$, $m$, and $F_{\rm fl}$ denote the position of nanoparticles, the noise in the feedback signal, the damping rate due to collisions with background gases, the damping rate due to an electric force, the oscillation frequency, the mass of trapped nanoparticles, and a stochastic force from background gases, respectively. 
%From the fluctuation-dissipation theorem, the stochastic force satisfies $\langle F_{fl}(t)F_{fl}^*(t') \rangle=2mk_{\rm B}T_0\Gamma_0 \delta(t-t')$ where the brackets, $k_{\rm B}$, and $T_0$ denote the expectation value, the Boltzmann constant, and the temperature of background gases, respectively.
%At weak feedback, the noise term originates from the measurement noise~\cite{poggio2007feedback}, while at strong feedback the noise term is dominated by sideband components in the feedback signal due to the oscillation outside the feedback bandwidth $\delta_{\rm BW}$. 
The PSD of the particle following eq.(\ref{eq:eom}) is given by~\cite{poggio2007feedback,gieseler2012subkelvin,vovrosh2017parametric}

\begin{align}
\label{eq:psd}
S(\Omega)&= \dfrac{2k_B T_0 \Gamma_0/m + \Omega^2 S_n(\Omega)}{(\Omega_0^2-\Omega^2)^2+\Omega^2 (\Gamma_0+\Gamma_{\rm c})^2}
%\dfrac{1}{2\pi}\int_{-\infty}^{\infty}\dfrac{\langle Q_n(\omega) Q_n^*(\omega')\rangle}{f(\omega)f^*(\omega')} d\omega'
\end{align}
where $k_{\rm B}$, $T_0$, and $S_n(\Omega)$ are the Boltzmann constant, the temperature of background gases, and the PSD of $q_n$, respectively. 

According to the kinetic theory for a particle in a gas, $\Gamma_0$ is proportional to the background pressure $P$ at our working pressure range~\cite{beresnev1990motion}:
\begin{align}
\label{eq:gamma0}
\Gamma_0 = B\dfrac{P}{R\rho}, \quad B=(4+\dfrac{\pi}{2})\sqrt{\dfrac{m_{\rm Air}}{2\pi k_{\rm B}T_0}}
\end{align}
where $\rho$, $m_{\rm Air}$, and $R$ are the density of the particle, the mass of background gas molecules, and the radius of the particle, respectively. Considering that the electric force is given as an amplitude of $neV/d_{\rm eff}$ multiplied by a sinusoidal time variation of $\dot{q}/(q_0 \Omega_0)$, where $n$ is the charge number, $e$ is the elementary charge, $V$ is the applied voltage amplitude, $d_{\rm eff}=V_{\rm cal}/E_{\rm cal}$ is the effective distance between electrodes calculated with a numerically simulated electric field $E_{\rm cal}$ for an applied voltage $V_{\rm cal}$ [Fig.~\ref{fig:expset}(b)], and $q_0$ is the amplitude of the motion at equilibrium, we find $\Gamma_{\rm c}=neV/(m\Omega_0 q_0 d_{\rm eff})$. 

%The PSD of the particle following eq.(\ref{eq:eom}) is given by~\cite{poggio2007feedback,gieseler2012subkelvin,vovrosh2017parametric}

%\begin{align}
%\label{eq:psd}
%S(\Omega)&= \dfrac{2k_B T_0 \Gamma_0/m + \Omega^2 S_n(\Omega)}{(\Omega_0^2-\Omega^2)^2+\Omega^2 (\Gamma_0+\Gamma_{\rm c})^2}
%\dfrac{1}{2\pi}\int_{-\infty}^{\infty}\dfrac{\langle Q_n(\omega) Q_n^*(\omega')\rangle}{f(\omega)f^*(\omega')} d\omega'
%\end{align}
%where $S_n(\Omega)$ is the PSD of $q_n$. 
%Eq.(\ref{eq:psd}) indicates that the spectral width $\Gamma_0+\Gamma_{\rm c}$ is broadened as $V$ increases. 
%Integrating eq.(\ref{eq:psd}) gives the variance of the position: 

%\begin{align}
%\label{eq:qvar}
%\langle q^2 \rangle=\dfrac{k_{\rm B}T_0}{m\Omega_0^2}\dfrac{\Gamma_0}{\Gamma_0+\Gamma_{\rm c}} 
%+\int_{0}^{\infty} \dfrac{\Omega^2 S_n(\Omega)/\pi d\Omega}{(\Omega_0^2-\Omega^2)^2+\Omega^2 (\Gamma_0+\Gamma_{\rm c})^2} 
%+\dfrac{S_{\rm n}}{2}\dfrac{\Gamma_{\rm c}^2}{\Gamma_0+\Gamma_{\rm c}}
%\end{align}

We first consider the case of weak feedback, where the influence of $q_n$ is negligible. Eq.(\ref{eq:psd}) then indicates that the PSD has the spectral width of $\Gamma_0+\Gamma_{\rm c}$, which is broadened with increasing $V$. 
%Eq.(\ref{eq:qvar}) is then a self-consistent equation for $q_0$ and provides the following representation for $T_{\rm eff}=m\Omega_0^2\langle q^2 \rangle/k_{\rm B}$: 
By integrating eq.(\ref{eq:psd}) to derive the variance of the position $\langle q^2 \rangle$, we obtain a self-consistent equation for $q_0$, resulting in the expression for $T_{\rm eff}=m\Omega_0^2\langle q^2 \rangle/k_{\rm B}$: 

%with $S_{\rm n}$ the PSD of the measurement noise. In the present study, the second term is much smaller than the first term. When the noise term is negligible, the effective motional temperature at equilibrium $T_{\rm eff}$ is expressed as~\cite{fnote7} 

\begin{eqnarray}
\label{eq:temp}
T_{\rm eff}=T_0\left(\sqrt{1+\alpha^2V^2}-\alpha V\right)^2, \quad
\alpha=\dfrac{ne}{2\Gamma_0d_{\rm eff}\sqrt{2mk_{\rm B}T_0}} 
\end{eqnarray}

Our model suggests that $T_{\rm eff}$ smoothly decreases from $T_0$ with increasing $V$. We experimentally confirm such a behavior by measuring $T_{\rm eff}$ for various voltage amplitudes at a fixed pressure [Fig.~\ref{fig:tvsv}(a)]. We realize cooling trapped nanoparticles in three dimensions by applying electric fields provided by three oscillators independently locked to the photodetector signal via a phase lock loop (PLL) [Fig.~\ref{fig:expset}(a)]~\cite{fnote7}. $T_{\rm eff}/T_0$ is obtained by comparing the areas of the PSDs around the oscillation frequency with and without cooling. In the present study, we assume $T_0=\unit[300]{K}$, which is expected to be a reasonable assumption at above $\unit[10]{Pa}$~\cite{hebestreit2018measuring}. For deriving the uncooled area, we use the PSD at around $\unit[10]{Pa}$ such that $T_0$ is not significantly deviate from $\unit[300]{K}$. The observed profile is in good agreement with the fitted curve based on our model, showing that our model provides a good understanding on the cooling process at weak feedback. 

Upon increasing $V$, however, we observe that $T_{\rm eff}$ deviates from eq.(\ref{eq:temp}) and increases with $V$. Experimentally we find that the minimum $T_{\rm eff}$ is obtained at around $\Gamma_0+\Gamma_{\rm c} \sim \delta_{\rm BW}$ with $\delta_{\rm BW} \approx 2\pi \times \unit[6]{kHz}$ being the bandwidth of PLL. Such a behavior is intuitively understood as follows: when the spectral width of the PSD exceeds $\delta_{\rm BW}$, the feedback loop starts to oscillate at around $\pm \delta_{\rm BW}$ and the feedback signal is dominated by undesirable frequency components amplifying the nanoparticle's motion [Fig.~\ref{fig:tvsv}(b),(c)]. At strong feedback, the feedback signal includes both cooling and heating components and the ratio between them depends not only on the feedback signal amplitude but also on the phase characteristics of the PLL, making it difficult to find the analytical representation of $T_{\rm eff}$ in the entire amplitude range. For understanding our results, we build a model based on our observation that the amplitude of the cooling component stays nearly constant $\Gamma_0+\Gamma_{\rm c} \approx \delta_{\rm BW}$ at strong feedback. 
%The first term of eq.(\ref{eq:psd}) is $k_{\rm B}T_0\Gamma_0/(m\Omega_0^2\delta_{\rm BW})$, while the second term of eq.(\ref{eq:psd}) is calculated by
With an assumption that the feedback signal is dominated by the two sideband components with equal amplitudes, we substitute $q_n=neV[\cos(\Omega_0t+\delta_{\rm BW}t)+\cos(\Omega_0t-\delta_{\rm BW}t)]/(\sqrt{2}m\Omega_0 d_{\rm eff})$ indicating off-resonant excitations~\cite{hebestreit2018calibration}. 
%Here we assumed that the feedback signal is dominated by the two sideband components with equal amplitudes. 
Considering $\delta_{\rm BW} \ll \Omega_0$, we find the representation of $T_{\rm eff}$ at strong feedback:

\begin{align}
\label{eq:strong}
T_{\rm eff} = T_0 \left[ \dfrac{\Gamma_0}{\delta_{\rm BW}}+\dfrac{(neV)^2}{10m\delta_{\rm BW}^2 d_{\rm eff}^2} \right]
\end{align}
which is in agreement with observed $T_{\rm eff}$ at large $V$ [Fig.~\ref{fig:tvsv}(a)].  

% increasing $V$ results in the increased width of the PSD 

 %an increase in $T_{\rm eff}$ with increasing the feedback voltage amplitude. Such a behavior occurs because as $\Gamma_{\rm c}$ is increased 

%the feedback loops oscillates at around the bandwidth of PLL $\delta_{\rm BW} \approx \unit[6]{kHz}$ and the feedback signal is dominated by undesirable frequency components amplifying the nanoparticle's motion. %Outside the bandwidth, the phase of the feedback signal is delayed by more than $\unit[180]{^\circ}$, amplifying the nanoparticle's motion instead of attenuating it. 
%At strong feedback, the feedback signal includes both cooling and heating components and the ratio between them depends not only on the feedback signal amplitude but also on the phase characteristics of the PLL, making it difficult to find analytical representations of $T_{\rm eff}$ in the entire amplitude range. For understanding our results and predicting the lowest $T_{\rm eff}$ at a given pressure, we build a model based on our observation that the amplitude of the cooling component stays nearly constant at $\Gamma_{\rm c} \geq \delta_{\rm BW}$. 

\begin{figure}[t]
\includegraphics[width=0.99\columnwidth] {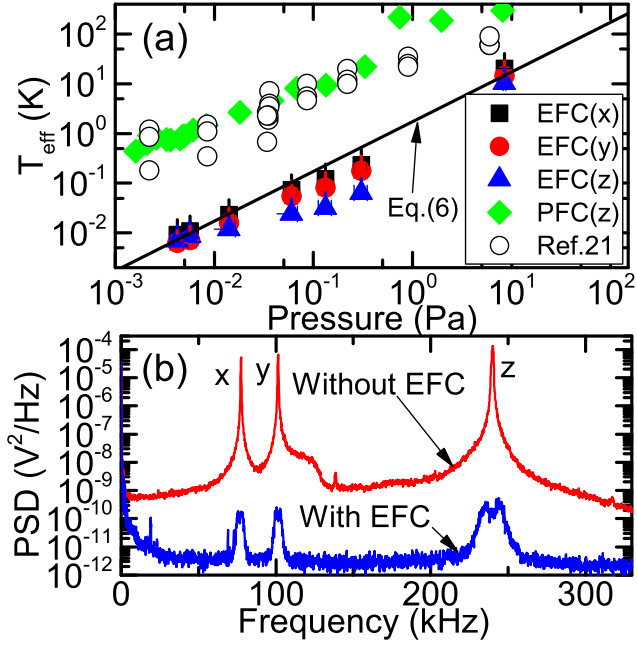}
\caption{(color online). (a) $T_{\rm eff}$ obtained with EFC as a function of the pressure. The error bars in $T_{\rm eff}$ are due to the systematic error of $T_0$ ($+\unit[300]{K}/-\unit[0]{K}$) estimated from ref.~\cite{hebestreit2018measuring}. Our results on PFC for another nanoparticle in the direction of the optical lattice is also presented. (b) The PSDs without EFC at $\unit[16]{Pa}$ and with EFC at $\unit[4\times10^{-3}]{Pa}$. }
\label{fig:tvsp}
\end{figure}
%Though increasing the PLL bandwidth is possible, this makes the PLL at low frequency inferior and also results in the increased $T_{\rm eff}$. 

%Finally, we address an important aspect regarding the limit of cooling. According to our model, $T_{\rm eff}$ should be infinitely decreased at infinitely large voltage amplitudes, which is opposed to our observation that the lowest achievable temperature is finite and varies at each pressure. 

At an optimum feedback amplitude, the minimum $T_{\rm eff}$ is approximately given by 
\begin{align}
\label{eq:minteff}
T_{\rm eff}^{\rm min}=T_0 \dfrac{\Gamma_0}{\delta_{\rm BW}}
\end{align}
suggesting that decreasing $P$ (and thus $\Gamma_0$) should result in lower $T_{\rm eff}$. By measuring $T_{\rm eff}$ in a wide pressure range between $\unit[0.004]{Pa}$ and $\unit[10]{Pa}$, we find that our simple model is in good agreement with experiments and observe a dramatic decrease in $T_{\rm eff}$ as $P$ is lowered (Fig.~\ref{fig:tvsp}). For comparison, we also show data points in previous work with PFC~\cite{gieseler2012subkelvin} and our results obtained with PFC. We find that within our working pressure range, the temperature obtained with EFC is lower than that with PFC by one to two orders of magnitude. The lowest observed temperature in the present work is between $\unit[6]{mK}$ and $\unit[10]{mK}$ at $\unit[4\times10^{-3}]{Pa}$ for all directions, with the phonon occupation numbers of $2.5\times 10^3$, $1.3\times 10^3$, and $6.1\times10^2$ in the $x$, $y$, and $z$ directions, respectively. 

We now discuss the limit of our approach at even lower pressures. From eq.(\ref{eq:minteff}), we anticipate that  $T_{\rm eff}$ reaches submillikelvin temperatures at $P < \unit[10^{-4}]{Pa}$. Recently, photon recoil heating has been identified as the main obstacle in cooling nanoparticles to submillikelvin temperatures with PFC~\cite{jain2016direct}. In the absence of feedback-induced heating, the phonon occupation number at low pressures is given by $\nu \sim \Gamma_{\rm rec}/(\Gamma_0+\Gamma_{\rm c})$, where $\Gamma_{\rm rec} \approx 2\pi \times \unit[10]{kHz}$ is a typical value for photon recoil heating~\cite{jain2016direct}. Hence, with the cooling capability of EFC demonstrated in the present study, photon recoil heating is not expected to be a major obstacle to cooling to near the ground state. Another issue which may prevent cooling at low pressures is the measurement noise affecting the feedback signal~\cite{poggio2007feedback}. Assuming that the feedback signal includes white noise originating from the noise in the position measurement, we take $q_n=\Gamma_{\rm c}q_{\rm mn}$ and find the representation of $T_{\rm eff}$ for $\Gamma_0 \ll \Gamma_{\rm c}$ as $T_{\rm eff}=T_0\Gamma_0/\Gamma_{\rm c}+m\Omega_0^2\Gamma_{\rm c}S_{\rm mn}/2k_{\rm B}$
%\begin{align}
%T_{\rm eff}=T_0\dfrac{\Gamma_0}{\Gamma_{\rm c}}+\dfrac{m\Omega_0^2\Gamma_{\rm c}}{2k_{\rm B}}S_{\rm mn}
%\end{align}
with $S_{\rm mn}$ the PSD of $q_{\rm mn}$. Using the typical value of $S_{\rm mn}=2\times \unit[10^{-25}]{m^2/Hz}$ for the $y$ direction estimated with the noise floor in our system, we find that the minimum value of $T_{\rm eff}=\sqrt{2m\Omega_0^2S_{\rm mn}T_0\Gamma_0/k_{\rm B}}$ is expected to be around $\unit[40]{\mu K}$ at $P=\unit[1 \times 10^{-6}]{Pa}$. Decreasing the pressure by two orders of magnitude, or placing the vacuum chamber in a cryogenic environment at $\unit[4]{K}$, both of which are feasible, will allow us to reach near the ground state ($\nu \sim 2$). If the detection efficiency, mainly limited by the numerical aperture of the lens, is improved, we expect that cooling to near the ground state will be possible at even higher pressures due to decreased $S_{\rm mn}$. Note that the noise floor for detecting the motion of nanoparticles is close to their ground state (several $\mu K$) in our setup.

%For the $z$ direction, 
%We find that at each pressure there exists an optimum voltage amplitude, above which $T_{\rm eff}$ increases (Fig.~\ref{fig:fbbw}). Such a limitation originates from the finite feedback bandwidth. When the voltage amplitude is increased, we observe that the feedback loops starts to oscillate at around $\unit[10]{kHz}$, which approximately coincides with the bandwidth of the phase lock loop (PLL) for locking the oscillators to the photodetector signals. Outside the bandwidth, the phase of the feedback signal is delayed by more than $\unit[180]{^\circ}$, amplifying the nanoparticle's motion instead of attenuating it. Though increasing the PLL bandwidth is possible, this makes the PLL at low frequency inferior and also results in the increased $T_{\rm eff}$. Thus, in the current setup, we have a limitation on the lowest achieved temperature originating from the PLL bandwidth. Note that this issue is not a fundamental problem in our cooling approach and can be overcome with a technical upgrade, e.g., to use filtered photodetector signals directly for feedback. 

%the feedback loop starts to oscillate and produces sidebands in the feedback signal. In such a case, 

%The effective distance between two electrodes for the $z$ direction is numerically determined to be $d_{\rm eff}^z=\unit[5.0(1)]{mm}$. 

\begin{figure}[t]
\includegraphics[width=0.99\columnwidth] {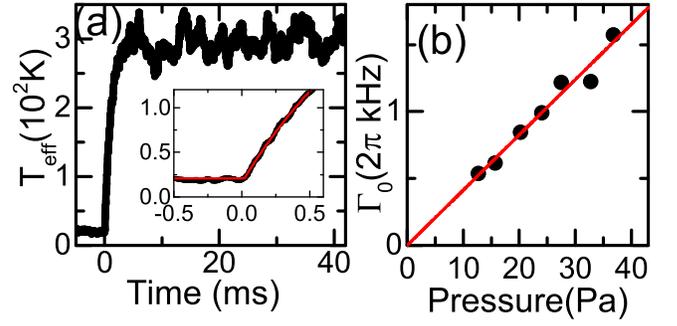}
\caption{ (a) Time evolution of $T_{\rm eff}$ in the $z$ direction averaged over 256 traces. EFC is turned off at $\unit[0]{ms}$. In the inset, we show the initial rise in $T_{\rm eff}$ with a fit, from which we derive $\Gamma_0$. (b) Measured $\Gamma_0$ with respect to the pressure. The solid line is a linear fit without intercept. The errors in $\Gamma_0$ are smaller than each point. The scatter of data points is mainly due to thermal fluctuations as observed in (a).}
\label{fig:retherm}
\end{figure}

In what follows, we show that EFC provides a unique means to measure the mass and the charge number of trapped nanoparticles that are crucial for understanding and predicting their behavior. The mass can be derived from $R$ obtained with eq.(\ref{eq:gamma0}) if $\Gamma_0$ is correctly measured. We find that the reliable way to determine $\Gamma_0$ is to observe the time evolution of $T_{\rm eff}$ in rethermalization experiments~\cite{hebestreit2018measuring}, instead of extracting the spectral width of the PSD that has been often used in previous work~\cite{vovrosh2017parametric}. We first prepare trapped nanoparticles cooled via EFC. After EFC is turned off, we observe an exponential growth of $T_{\rm eff}$ [Fig.~\ref{fig:retherm}(a)]. The time constant of the growth is the inverse of $\Gamma_0$. 
%In this way, the influence of the thermal fluctuation of the oscillation amplitude is minimized. 
The spectral width extracted from the PSD tends to exhibit broadening by fluctuations in the oscillation frequency presumably due to the vibration of optics, the fluctuations in laser intensity, and thermal nonlinearities~\cite{gieseler2013thermal,yoneda2017thermal}. We confirm that at our working pressure range the measured values of $\Gamma_0$ are proportional to the pressure as expected from eq.(\ref{eq:gamma0}) [Fig.~\ref{fig:retherm}(b)]. The slope of this plot $g$ is used for deriving $m=4\pi B^3/(3\rho^2g^3)$. 

\begin{figure}[t]
\includegraphics[width=0.95\columnwidth] {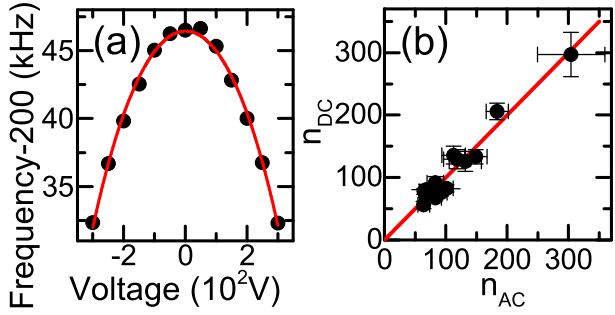}
\caption{(color online). (a) The oscillation frequency in the $z$ direction as a function of the applied dc voltage.  A fit with eq.(\ref{eq:freqgrad}) is shown by a solid line.  (b) Comparison of measured charge numbers, $n_{\rm AC}$ and  $n_{\rm DC}$, for different nanoparticles trapped during multiple experimental runs. The error bars mainly originate from the errors in determining $g$. The solid line shows $n_{\rm AC} = n_{\rm DC}$ and is not a fit. }
\label{fig:ndc}
\end{figure}

The value of $\alpha$ obtained by fitting eq.(\ref{eq:temp}) to the cooling curve at weak feedback [Fig.~\ref{fig:tvsv}(a)] reveals the charge number as $n=4d_{\rm eff}\alpha P\sqrt{2\pi k_{\rm B}T_0B^3/3g}/\rho e$, which we denote as $n_{\rm AC}$. For the $z$ direction, we have $d_{\rm eff}=\unit[5.0]{mm}$. To test whether the obtained values of $n_{\rm AC}$ are correct, we introduce a new method for determining $n$ independently from the measurement of Fig.~\ref{fig:tvsv}(a). We apply a dc electric field and observe an induced shift in the oscillation frequency in the $z$ direction. The presence of a gradient shifts the position of the potential minimum, at which the oscillation frequency is lowered due to the anharmonicity of the potential. By measuring the frequency shift in the direction of the optical lattice, which has a well-defined sinusoidal structure with a spacing of $\lambda/2$, we obtain the applied gradient precisely. The oscillation frequency in the $z$ direction in the presence of a dc voltage $V$ is given by~\cite{fnote7}
\begin{align}
\label{eq:freqgrad}
%\Omega_z=\Omega_{z0}\left[1-\left\{\dfrac{4\pi (neV/d_{\rm eff}+F_{\rm 0})}{m\Omega_{z0}^2\lambda}\right\}^2\right]^{1/4} \\
\Omega_z=\Omega_{z0}\left[1-\left\{\dfrac{CV+F_0/(\pi m \lambda)}{(\Omega_{z0}/2\pi)^2}\right\}^2\right]^{1/4},
C=\dfrac{ne}{\pi m \lambda d_{\rm eff}}
\end{align}
where $\Omega_{z0}$ is the oscillation frequency in the absence of gradients and $F_{\rm 0}$ is the offset gradient mainly due to the radiation pressure by the trapping laser. We observe that the oscillation frequency varies as a dc electric field is applied, in good agreement with eq.(\ref{eq:freqgrad}) [Fig.~\ref{fig:ndc}(a)]. From the curvature of this plot $C$, we obtain the value of $n=4\pi^2 \lambda d_{\rm eff}CB^3/(3\rho^2g^3 e)$, which we denote as $n_{\rm DC}$. Here we find that the two methods for determining $n$ have different dependences on $g$: $n_{\rm AC}\propto g^{-1/2}$ and $n_{\rm DC}\propto g^{-3}$. This fact indicates that unless the value of $g$ is properly measured we encounter an unrealistic result of $n_{\rm AC}\neq n_{\rm DC}$.  In fact, with the conventional method for determining $g$, i.e., extracting the spectral width from the PSD~\cite{gieseler2012subkelvin,vovrosh2017parametric}, we found large discrepancies between $n_{\rm AC}$ and $n_{\rm DC}$ by up to a factor of 10. 

With $g$ obtained with the rethermalization measurements, we arrive at reliable values of $n$ with $n_{\rm AC}$ and $n_{\rm DC}$ agreeing within $\unit[25]{\%}$. For the specific case of the presented data in Fig.~\ref{fig:retherm} and Fig.~\ref{fig:ndc}(a), the mass and the charge number are $m=\unit[5.3(3)\times10^{-18}]{kg}$ and $n_{\rm DC}=74(5)$, respectively, where the errors are dominated by the error in determining $g$. The comparison between $n_{\rm AC}$ and $n_{\rm DC}$ for multiple experimental runs is shown in Fig.~\ref{fig:ndc}(b). In our system, we have not succeeded in observing the variation of $n$ by a single $e$, which was used to determine $n$ in previous work~\cite{frimmer2017controlling}. We infer that this is because the charge variation in our system is much larger than in the previous work. The large charge variation possibly suggests that ionized gas molecules around the trap region are denser than in previous work.

In conclusion, we demonstrate an efficient approach for cooling the center-of-mass motion of single optically trapped nanoparticles via the combination of the optical observation of the particle position and the application of electric fields synchronized to their motion. Our cooling approach provides a unique means to characterize the properties of trapped nanoparticles. The demonstrated temperatures of below $\unit[10]{mK}$ at $\unit[4\times10^{-3}]{Pa}$ are about one to two orders of magnitude lower than those obtained with the conventional method at this pressure. The advantages of our approaches lie in the strong cooling force independent from $T_{\rm eff}$ and the clean observation signals unaffected by feedback signals. We envision that cooling to near the ground state is achieved at a pressure of $1\times \unit[10^{-8}]{Pa}$, or at a pressure of $1\times \unit[10^{-6}]{Pa}$ in a cryogenic environment at $\unit[4]{K}$. Due to the high cooling rate of our approach, photon recoil heating, which has been a severe limitation with the conventional optical cooling, will not be a major obstacle down to around the ground state; rather, the noise in the feedback signal originating from position measurements will be of the largest concern. Our cooling strategy is also applicable to particles trapped in an ion trap, where they can be observed with a weak probe laser that has a minimum heating effect. %Our approach allows us to prepare cold nanoparticles for various measurements such as calibration of the PSD and force sensing even at medium vacuum.

{\it Note added}: After the submission of this manuscript, two related works by the group of ETH Zurich~\cite{tebbenjohanns2018cold} and by the group of ICFO~\cite{conangla2018optimal} have appeared on arXiv.

%Acknowledgment:\\
\begin{acknowledgments}
We thank M.\,Ueda and M.\,Kozuma for fruitful discussions. This work is supported by the Murata Science Foundation, the Mitsubishi Foundation, the Challenging Research Award and the 'Planting Seeds for Research' program from Tokyo Institute of Technology, Research Foundation for Opto-Science and Technology, JSPS KAKENHI (Grants No. JP16K13857 and JP16H06016), and JST PRESTO (Grant No. JPMJPR1661).
\end{acknowledgments}

\section{Supplementary Information}
\subsection{Experimental setup}
A single-frequency infrared laser at a wavelength of $\unit[1550]{nm}$ with a power of about $\unit[400]{mW}$ is focused to a beam waist of about $\unit[1.5]{\mu m}$ with an objective lens (NA=0.85) and is retro-reflected to form a standing-wave optical trap (an optical lattice). The setup around the trap region is installed in a vacuum chamber. A piezo module attached to the retro-reflecting mirror allows us to precisely control the position of the trapped nanoparticles in the $z$ direction. The light is linearly polarized along the $y$ direction. The intensity of the trapping beam is controlled by an acousto-optic modulator (AOM). A part of the trapping beam is extracted at a beam splitter and is used for detecting the motion of nanoparticles via a home-made balanced photodetector, which subtracts the signal without nanoparticles from the signal with nanoparticles. The subtraction is to minimize the influence of the laser intensity noise on the signal. The nanoparticle's motion in the $z$ direction modulates the total intensity of the trapping beam, while those in the $x,y$ directions modulate the spatial beam profile. The light from the trapping beam is incident on the photodetector with a slight misalignment in both $x$ and $y$ directions such that the spatial intensity modulation is also detected. In this way, we observe the three dimensional motion of a trapped nanoparticle with a single photodetector. 

We generate the feedback signal with a digital lock-in amplifier (MFLI, Zurich Instruments). The oscillator incorporated in the lock-in amplifier is locked to the photodetector signal with a phase lock loop (PLL). A phase difference of $\unit[90]{^\circ}$ between the photodetector signal and the oscillator output signal is introduced at PLL. The output signal of the oscillator is then sent to the electrodes installed in the vacuum chamber. The signal for the $z$ direction is applied to two objective lens with metallic housing, while the signal for the $x$ and $y$ directions is applied to two tungsten wires perpendicular to the trapping beam [see Fig.~1(a)]. We perform a finite element matrix calculation of electric fields around the trapping region. The configuration of the electrodes and the lenses and the calculated electric fields in the $z$ direction $E_z$ on a plane $\unit[45]{^\circ}$ tilted with respect to the $xz$ plane are shown in Fig.~1(b). 

\subsection{Fitting on rethermalization curves}
We find that extracting the exponential time constants $1/\Gamma_0$ from the rethermalization curves [Fig.~4(a)] requires a special attention. When we fitted the exponential function in the entire time range from $t<0$ to $t\gg 1/\Gamma_0$, we observed large fluctuations in $\Gamma_0$ because of the thermal fluctuation of the equilibrium temperatures at $t\gg 1/\Gamma_0$. To remove the influence of thermal fluctuations and derive the proper value of $g=\Gamma_0/P$, we take the following procedure. First, measure the rethermalization curves for $N(\approx10)$ pressure values. Second, for each curve we take the average values of the equilibrium temperatures ($T_i, i=1,2,\cdots N$) at $t\gg 1/\Gamma_0$. Third, we define the equilibrium temperatures $T_{eq}$ as a mean of the values of $T_i$. Finally, using $T_{eq}$ as a fixed parameter, we fit the curve only at around the initial rise and derive $\Gamma_0$. 

\subsection{Derivation of the shift in the oscillation frequency induced by a dc electric field}
The optical lattice potential is given by
\begin{align}
\label{eq:lat}
U_1(z)=\dfrac{U_0}{2}\left[1-\cos\left(\dfrac{4\pi z}{\lambda}\right)\right]
\end{align}
with $U_0$ the potential depth. The oscillation frequency without a gradient $\Omega_{z0}$ is related to $U_0$ by 
\begin{align}
\label{eq:latfreq}
U_0=\dfrac{m\Omega_{z0}^2\lambda^2}{8\pi^2}
\end{align}
The gradient from the DC electric field and the radiation pressure $F_{\rm 0}$ is
\begin{align}
U_2(z)=\left(neV/d_{\rm eff}+F_{\rm 0}\right)z
\end{align}
The minimum of $U_{\rm tot}(z)=U_1(z)+U_2(z)$ appears at $z=z_0$ satisfying $dU_{\rm tot}(z_0)/dz=0$, which is written as
\begin{align}
\label{eq:minz0}
\sin\left(\dfrac{4\pi z_0}{\lambda}\right)=-\dfrac{\lambda\left(neV/d_{\rm eff}+F_{\rm 0}\right)}{2\pi U_0}
\end{align}
The oscillation frequency around $z=z_0$ can be obtained by calculating the second derivative of $U_{\rm tot}(z)$:
\begin{align}
\Omega_z^2=& \dfrac{1}{m}\dfrac{d^2U_{\rm tot}}{dz^2} \notag \\
=& \dfrac{8\pi^2 U_0}{m\lambda^2}\cos\left(\dfrac{4\pi z_0}{\lambda}\right)
\end{align}
Substituting the expressions (\ref{eq:latfreq}) and (\ref{eq:minz0}), we obtain a representation for $\Omega_z$ as shown in eq.(7) in the main text.

\bibliographystyle{apsrev}

%\bibliography{NPbib,ultracold,ErBEC}

\begin{thebibliography}{39}
\bibitem[{fno({\natexlab{b}})}]{fnote8}
\bibinfo{note}{The chamber is equipped also with a capacitance gage, providing
  the pressure values between 0.5 and $\unit[133]{Pa}$ with an accuracy of
  $\unit[0.27]{Pa}$. Outside this pressure range, the pressure values are
  recorded with an accuracy of $\unit[15]{\%}$.}
\end{thebibliography}

\end{document}